\documentclass{iau}
\usepackage{graphicx}
\usepackage{amsmath}
\title[S340.~~Standard Magnetic Field Production at HSOS] %% give here short title %%
{Standard Magnetic Field Production at Huairou Solar Observing Station}

\author[Liu et al.]   %% give here short author list %%
{Suo Liu$^{1,2}$ \& Ganghua Lin$^{1,2}$ \& Xiao Yang$^{1,2}$ \& Xiaofan Wang$^{1,2}$ \& Jiangtao Su$^{1,2}$ \& Yuanyong Deng$^{1,2}$}

\affiliation{$^1$Key Laboratory of Solar Activity, National Astronomical Observatories, Chinese Academy of Sciences \\ email: {\tt lius@nao.cas.cn} \\[\affilskip]
$^2$School of Astronomy and Space Science, University of Chinese Academy of Sciences}

\pubyear{2018}
\volume{340}  %% insert here IAU Symposium No.
\jname{Long-term datasets for the understanding of solar and stellar magnetic cycles}
\editors{Dipankar Banerjee, Sami Solanki \& Jie Jiang, eds.}
\begin{document}

\maketitle

\begin{abstract}
 The regular solar observations are operated at Huairou Solar Observing Station (HSOS) since 1987, which make the construction of long-term magnetic field datasets available to understand solar magnetic field and cycles. There exist some inconveniences for solar physicist to use these data, because the data storage medium and format at HSOS experienced some changes. Additionally, the processes of magnetic field calibration are not easy to deal with for who are not familiar with these data. Here shows that the magnetic field of HSOS are further processed toward international standards, in order to explore HSOS observations data for scientific research.

\end{abstract}

\section{Introduction}
Solar magnetic field is the most important observable measurement, which can contribute to the solar activities and the understanding of the nature of solar cycles, especially for long time observation of solar magnetic field. However, some instruments operated for many years, such as Solar Magnetic Field Telescope (SMFT), there are the inconsistencies of observations data during their operated times. Thus, it is an urgent task to make various observation data consistent and satisfying international standard, which include the format of data and the data information (keywords about information).
Flexible Image Transport System (FITS: http://fits.gsfc.nasa.gov/) format be regarded as a standard astronomical data production at present, thus FITS should be selected for standard data format for data production.
As for magnetic field obtained from SMFT, a filter-style instrument, the processes of magnetic field calibration is needed, since the Stokes (I, Q, U, V) observed directly should convert to magnetic field, which are deduced and expected physical quantity for scientists to use conveniently and easily.
This paper will introduce simply the magnetic field observations by SMFT, and shows that magnetic field process toward standard data production
for solar physicists to use.

\section{Observation Data and Calibration}
The routine solar magnetic field of photosphere observations are available at HSOS since 1987, the magnetic field observations are carried out by SMFT that is a 35-cm vacuum refraction solar telescope (Ai \& Hu, 1986). The birefringent filter work at FeI 5324.19 \AA~ for photosphere vector magnetic field, then it is reconstructed from Stokes parameters (I, Q, U and V).  The linear calibration is used to calibrate magnetic field under the weak-field approximation.
$~B_{L}=C_{L}V,~B_{T}=C_{T}(Q^{2}+U^{2})^{1/4},~\theta=arctan(\dfrac{B_{L}}{B_{\bot}}),~\phi=\dfrac{1}{2}arctan(\dfrac{U}{Q})$,
where B$_L$ and C$_L$ are the line-of-sight (LOS) magnetic field and the calibration coefficients, B$_T$ and C$_T$ are transverse magnetic field and its coefficients. $\theta$ and $\phi$ are the inclination and azimuth of field (Su \& Zhang, 2004).

For SMFT, the storage medium, read methods and format of data have been changed many times during its operated periods.
Data storage medium has undergone the tape, optical disk, CD-ROM and hard disk with the time flowing.
At present, the observations data is stored in the form of electronic one, however the data format conversion from others to FITS should be done in this work.

\begin{figure}[b]
 \vspace*{-.0 cm}
\begin{center}
 \includegraphics[width=4.7in]{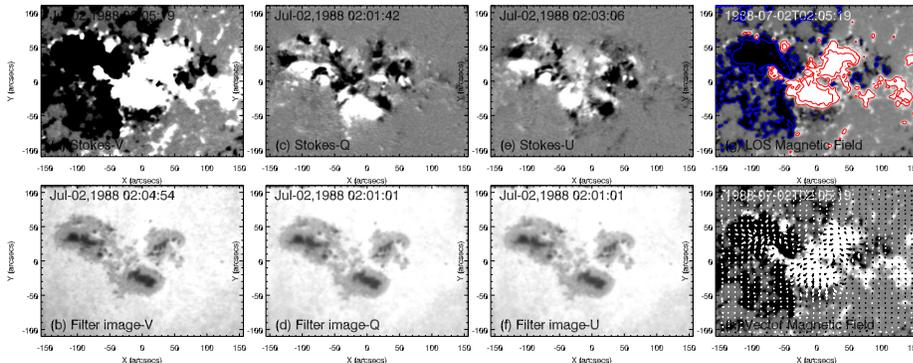}
% \vspace*{-1.0 cm}
 \caption{(a)-(f): the observed image of Stokes-V/Q/U and the corresponding filer image. (g): the LOS magnetic field with its contours.
 (h): the vector magnetic field, gray scale is light of sight field and the arrows indicate transverse field }\label{Fig1}
   \label{fig1}
\end{center}
\end{figure}
\section{Data Processing and Conclusions}

For SMFT, a group stokes Q/U/V are used to deduce vector magnetic field,
which is the process that contain magnetic field calibration and spatial alignment of magnetic field components.
For each Stokes (/Q/U/V), the corresponding filter images are simultaneously observed,
which are used to compensate for the time differences during the observations of Q, U and V.
The reconstruction of vector magnetic field are as follows:
1) Find the group of Q/U/V (time difference less than 15 minutes) and their filter image.
2) Calculate the position of compensation of Q/U/V basing on its individual filter image.
3) The weak-field approximation is use to calibrate magnetic field.
4) Then the transverse field ($B_{t}$) is decomposed to two components on the surface indicated by $B_{x}$ and $B_{y}$.
5) Write magnetic field indicating by $B_{z}$, $B_{x}$ and $B_{y}$ as FITS and make the data information (header) more perfect.
The figure \ref{Fig1} show this process, (a)-(f) give the original observation, and the expected physical quantities are exhibited in (g) and (h).
Recently, the magnetic field more than 30 years (1987-2018) observed by SMFT are processed to produce standard productions and can be found at http://sun.bao.ac.cn.

Grants: 2014FY120300, U1531247, XDB09040200, 11473039, 11373040, 11673033,11203036.


\begin{thebibliography}{}

\bibitem[Ai \& Hu {1986}]{1986AcASn..27..173A}
{Ai, G.X.~\& Hu, Y.F.,}
1986, \textit{Acta Astronomica Sinica}, 27, 173

\bibitem[Su \& Zhang {2004}]{2004ChJAA...4..365S}
{Su, J.T.~\& Zhang, H.Q.,}
 2004, \textit{Chin. J. Astron. Astrophys.}, 4, 365



\end{thebibliography}
\end{document}